# ELECTRIC FIELD GRADIENTS IN MgB$_2$ SYNTHESIZED AT HIGH PRESSURE: $^{111}$Cd TDPAC STUDY AND *AB INITIO* CALCULATION

A.V. Tsvyashchenko[a,*], L.N. Fomicheva[a], M.V. Magnitskaya[a], E.N. Shirani[a],
V.B. Brudanin[b], D.V. Filossofov[b], O.I. Kochetov[b], N.A. Lebedev[b], A.F. Novgorodov[b]
A.V. Salamatin[b], N.A. Korolev[b], A.I. Velichkov[b], V.V. Timkin[b],
A.P. Menushenkov[c], A.V. Kuznetsov[c], V.M. Shabanov[c] and Z.Z. Akselrod[d]

[a] Institute for High Pressure Physics, 142190 Troitsk, Moscow Region, Russia
[b] Joint Institute for Nuclear Research, 141980 Dubna, Moscow Region, Russia
[c] Moscow Engineering-Physics Institute, 115409 Kashirskoe shosse 31, Moscow, Russia
[d] Institute of Nuclear Physics, Moscow State University, 119899 Vorob'evy gory, Moscow, Russia

**Abstract**

We report the high-pressure synthesis of novel superconductor MgB$_2$ and some related compounds. The superconducting transition temperature of our samples of MgB$_2$ is found to be 36.6 K. The MgB$_2$ lattice parameters determined via X-ray diffraction are in excellent agreement with results of our *ab initio* calculations. The TDPAC measurements of $^{111}$Cd quadrupole frequency ***n***$_Q$ demonstrate a small increase in ***n***$_Q$ with decreasing temperature from $T_{room}$ to $T_{He}$. The electric field gradient $V_{zz}$ at the B site calculated from first principles is in fair agreement with $V_{zz}$ obtained from $^{11}$B NMR spectra of MgB$_2$ reported in the literature. It is also very close to $V_{zz}$ found in our $^{111}$Cd TDPAC experiments, which suggests that the $^{111}$Cd probe substitutes for boron in the MgB$_2$ lattice.

## 1. INTRODUCTION

The recent discovery of intermediate-temperature superconductivity in magnesium diboride MgB$_2$ with $T_c \sim 39$ K[1] and its almost simultaneous explanation in terms of phonon-mediated BCS superconductivity based on the observation of significant boron isotope effect[2] has triggered an avalanche of theoretical work including *ab initio* electronic band structure calculations (see, e.g.[3,4,5,6]). A recent $^{11}$B NMR study[7] has revealed an exponential temperature dependence of the nuclear relaxation rate $1/T_1$ in the superconducting (SC) state, with a relatively large SC gap of $2\mathbf{D}/k_B T_c \sim 5$ implying strong-coupling s-wave superconductivity in this system. Other $^{11}$B NMR studies[8,9] of MgB$_2$ have provided important information on the local electronic state of the boron site by measuring the [11]B quadrupole coupling constant $n_Q = eQV_{zz}/h$ that can be evaluated theoretically from first-principles calculations.

---

[*] Fax: (7-095) 334-0012; E-mail: tsvyash@ns.hppi.troitsk.ru

In NMR experiments, however, the resonance frequency of [11]B is determined both by the hyperfine magnetic interaction and by the hyperfine interaction of the [11]B quadrupole moment $Q$ with electric field gradient (EFG) $V_{zz}$, created at the boron site by electronic and ionic environment. This combined interaction complicates the determination of $n_Q$. For instance, Tou *et al.*[10] have found that in the normal state ($T = 50$ K) $n_Q$ obtained from magnetic-field dependence of the NMR frequency shift is equal to 1100±150 kHz. However, $n_Q$(FWHM) determined through the relation $n_Q$(FWHM) = $1.3856(n_0\Delta n)^{1/2}$, where $\Delta n$ is the width between the peak and shoulder of NMR spectrum, is equal to 835 kHz[10], in close agreement with the $n_Q$ values found in other NMR experiments[8,9] (see Table 2 below).

In this paper, the results of high-pressure synthesis of $MgB_2$, as well as of some related compounds are presented. We use the time-differential perturbed angular γγ-correlation (TDPAC) method to investigate the hyperfine quadrupole interaction at [111]Cd probe nucleus in the $MgB_2$ lattice consisting of alternating graphite-like layers of B atoms and hexagonal layers of Mg atoms, with axial symmetry of electric field gradients. Previously, this high-pressure technique has been successfully applied, e.g., for synthesis of high-pressure phases of $YbFe_2$ with subsequent [111]Cd TDPAC investigation of these phases[11]. In contrast to the NMR experiments, this PAC method allows to determine $n_Q$ in $MgB_2$ directly, since there is no hyperfine magnetic interaction on the [111]Cd nucleus, owing to the lack of magnetic ordering in $MgB_2$. We present also *ab initio* band-structure calculations of EFG's at the B and Mg atoms in $MgB_2$ and compare these results with our experimental data.

## 2. HIGH-PRESSURE SYNTHESIS

Magnesium diboride samples were prepared by reacting stoichiometric amounts of powdered B and Mg at constant temperature $T \geq 900°C$ and constant pressure of 6 ~ 8 GPa in a high-pressure "Toroid" cell constructed by Khvostantsev *et al.*[12]. Two different experimental techniques were applied. At first, the pellet was prepared by compacting the well-mixed powdered constituents at room temperature. With the first technique of solid-phase synthesis, the pellet was placed into the tantalum heater where it was held for 20 – 40 min at the constant pressure of 6.2 GPa and temperature of 950°C which is lower than $T_{melt}$ of Mg at this pressure[13]. In this case, temperature was controlled by a chromel-alumel thermocouple located outside the heater near its wall. In the second case, the pellet was placed in a rock-salt pipe ampoule and then directly heated electrically to $T \sim 1500°C$ (above $T_{melt}$ of Mg) at a constant pressure in the range $6 < p < 8$ GPa. In the both cases the experimental sample was then rapidly quenched to room temperature, and pressure was released. The structural properties of the samples obtained with both techniques were found to be identical, and their SC properties are also quite similar.

After high pressure–high temperature treatment, the samples were investigated at normal conditions. The structure and phase composition of the samples obtained were determined from X-ray powder patterns with a 114-mm camera RKU-114 and Cu $K_\alpha$ radiation. X-ray analysis showed that at ambient conditions the $MgB_2$ samples have the hexagonal $AlB_2$-type structure, space group P6/mmm, and lattice parameters $a = 3.075$Å, $c = 3.519$Å, $c/a = 1.144$. A weak trace of MgO was also observed in the powder patterns. AC susceptibility measurements showed the SC transition with $T_c = 36.6$ K. The results of these measurements are presented in Fig. 1.

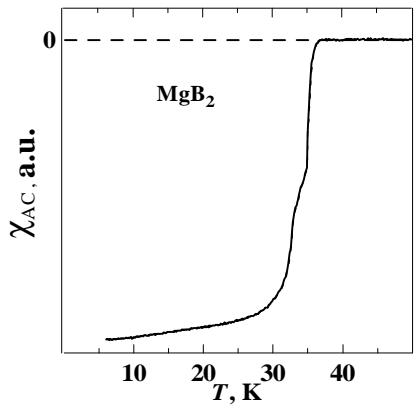

Fig.1. Magnetic susceptibility of $MgB_2$ as a function of temperature.

It should be mentioned that we synthesized also the compounds $BeB_2$, $CaB_2$ and $Ca_{0.5}Mg_{0.5}B_2$ using the same high-pressure technique. All these three alloys do not belong to the $AlB_2$-type structure and are not superconducting up to 4.2 K. Our results on $BeB_2$ agree with other investigations of this compound[14].

Besides, we synthesized compounds in the systems $Mg(B_{1-x}Ga_x)_2$ at $0.05 \leq x \leq 1.0$ and $Mg_{1-x}Sc_xB_2$ at $0 \leq x \leq 0.3$. The substitution of Mg with Sc was found to reduce $T_c$ considerably, but the $AlB_2$-type crystal structure remains unchanged. Similar results have been obtained by other groups investigating the substitution of Mg with Al[15] and Li[16]. In turn, the replacement of B with Ga reduces $T_c$ as well, which correlates with the results of other experiments[17] on boron sublattice doping. In this series of experiments, however, a single phase of composition close to MgBGa was synthesized, which possessed $T_c = 32$ K. At normal conditions this high-pressure phase is rather unstable — it breaks down to the mixture of powdered $MgB_2$ and $MgGa_2$ in a few days. The crystal structure of MgBGa was tentatively determined as cubic $SrSi_2$-type structure (perhaps, distorted) with lattice parameter $a = 6.291$Å. In this context, we would like to mention the recent work by Sanfilippo et al.[18] who observed superconductivity with $T_c = 14$ K in high-pressure phase of $CaSi_2$ ($AlB_2$-type structure) at $p = 16$ GPa. It should be noted that various low-pressure and high-pressure phases of both disilicides, $CaSi_2$ and $SrSi_2$, are characterized by 3-connected nets of Si[19].

## 3. TDPAC MEASUREMENTS

$^{111}$In was produced via the reaction $^{109}$Ag{α, 2n}$^{111}$In by irradiation of a metallic silver target with α particles ($Å_\alpha$ = 30 MeV) at the U-200 cyclotron of JINR, Dubna. $^{111}$In$^{20}$ of high specific activity was diffused into the MgB$_2$ samples at 550°C for 2.5 hrs. The thermal diffusion proceeded in vacuum to prevent oxidation, and a tantalum sample holder was used. The excited state (spin $I$ = 7/2$^+$, energy $E$ = 420 keV) of $^{111}$Cd populated via the electronic-capture decay of $^{111}$In de-excites, by way of the γγ-cascade of 171–245 keV, to the ground state ($I$ = 1/2$^+$) through the isomeric intermediate state of $^{111}$Cd with $I$ = 5/2$^+$, $E$ = 247 keV, half-period $t_{1/2}$ = 85 ns and nuclear quadrupole moment $Q$= +0.83(13) b$^{21}$. The angular correlation of this cascade is largely anisotropic ($Å_2^{max}$ = –18.0%).

A four-detector TDPAC spectrometer, with detectors arranged in a plane at 90° angular intervals and controlled by a personal computer has been developed[22] for detecting PAC at different temperatures and pressures. The four NaI(Tl) cylindrical detectors (40x40 mm) coupled to the Philips XP2020Q photomultipliers provide the time resolution of 2.5 ns (FWHM) per detector pair for γγ-cascade of 172–247 keV of $^{111}$Cd. The low-temperature experiments can be performed with a nitrogen cryostat.

The components of the EFG tensor referred to its principal axes are denoted as $V_{xx}$, $V_{yy}$ and $V_{zz}$. In the case of hexagonal MgB$_2$ lattice, the coordinate axes are chosen such that $V_{zz}$ is the EFG component along the crystallographic $c$ axis and $|V_{zz}| > |V_{yy}| \geq |V_{xx}|$. Then the EFG tensor is completely characterized by two parameters: the largest component $V_{zz}$ and the axial asymmetry parameter **h** defined as **h** = $|V_{xx} - V_{yy}|/|V_{zz}|$.

Data processing aimed at the determination of these parameters starts with calculation of a ratio $R(t)$ from the experimental spectra $N(J, t)$:

$$R(t) = -2[N(180°, t) - N(90°, t)]/[N(180°, t) + 2N(90°, t)] \ .$$

For polycrystalline samples, with the assumption of a Lorentzian distribution of EFG, theoretical calculations[23,24] give

$$R(t) = Å_{22}^{eff} \sum_n s_{2,n} \cos(a_n w_0 t) \exp(-a_n L t) \ ,$$

where $w_0$ is the quadrupole frequency, $w_0 = 3\pi |eQV_{zz}/h|/I(2I-1)$ for half-integral spin $I$. The frequency $w_0$ is a function of the quadrupole coupling constant $n_Q = eQV_{zz}/h$. Here $Q$ is the nuclear quadrupole moment for $^{111}$Cd intermediate state of γ-ray cascade, $h$ is the Planck's constant and $L$ is a relative width of the quadrupole frequency distribution due to lattice defects and other imperfections. The coefficients $a_n$ and $s_{2,n}$ depend on $n$ and $h$[24]. For $h = 0$, $a_n$ is equal to $n$, resulting in a periodic $R$-vs.-$t$ pattern.

In Figure 2, the TDPAC spectra taken both at room temperature (a) and at $T = 143$ K (b) are displayed. The least-squares fits of the theoretical relation $R(t)$ to the experimental data were obtained by means of an appropriate MATLAB routine.

The spectrum measured at room temperature is characterized by a nonperiodic perturbation corresponding to a rather broad distribution of the quadrupole frequency with an average value of $n_Q$ equal to 31.0(1) MHz and $L = 0.12$, which gives $V_{zz} = 15.4\times10^{16}$ V/cm$^2$. The spectrum taken at $T = 143$ K is characterized by a periodic perturbation corresponding to a smaller distribution of the quadrupole frequency, with the average equal to 37.5(1) MHz and $L = 0.05$, which gives $V_{zz} = 18.6\times10^{16}$ V/cm$^2$.

Thus, the quadrupole coupling constant $n_Q$ (and, correspondingly, $V_{zz}$) decreases with increasing temperature. This result does not contradict to the well-known law of $T^{3/2}$ applicable mainly to the probe nuclei of sp elements[23], $n_Q(T) = n_Q(0)(1 - BT^{3/2})$, where $n_Q(0)$ is the quadrupole coupling constant at 0 K. For MgB$_2$, the value of parameter $B$ obtained by fitting the $n_Q(T)$ law to the experimental data is $B \sim 4.8\times10^{-5}$ K$^{-3/2}$. This formula extrapolated from the normal state to lower temperatures gives an estimate $n_Q(0) \sim 40.9$ MHz, corresponding to $V_{zz} \sim 20.3 \times10^{16}$ V/cm$^2$.

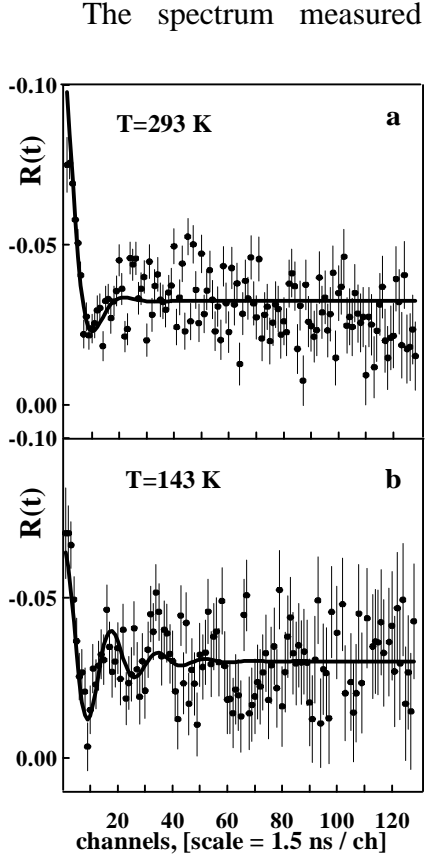

Fig. 2. $^{111}$Cd TDPAC spectra for MgB$_2$ at two temperatures. The solid curves are the fits of $R(t)$ to the experimental data.

4. *AB INITIO* CALCULATIONS

We calculated the electronic band structure and ground-state (i.e. at $T = 0$ K) properties of MgB$_2$ within density functional theory[25] using the full-potential linearized augmented plane wave (FPLAPW) method as implemented in the WIEN97 code[26], with the generalized gradient approximation (GGA)[27] for the exchange-correlation potential. The muffin-tin radii of Mg and B equal to 1.8 and 1.5 a.u., respectively, were kept constant on varying both the volume and $c/a$. The mesh of 162 **k**-points was used in the irreducible wedge of the Brillouin zone. The Mg 2s and 2p states were treated as band states using the local orbital extension of the LAPW method[28]. Parameters of our calculation were as follows: the potential and the charge density were expanded up to $l_{max} = 10$ and $g_{max} = 15$ a.u.$^{-1}$, the cutoff $r_{MT}k_{max}$ was chosen to be equal to 8.0.

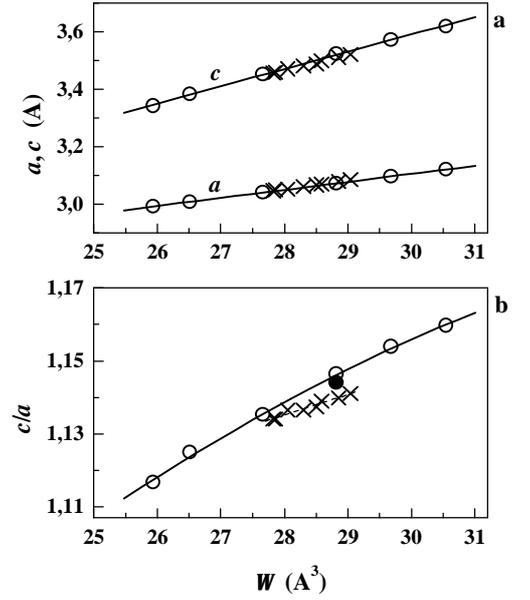

Total energy was calculated as a function of unit cell volume $W$ over the range $(0.88 - 1.09)W_{exp}$, where $W_{exp}$ is the experimental equilibrium volume at normal pressure. At each value of $W$, the $c/a$ ratio was optimized. An energy-vs.-volume relation obtained by integrating the Birch $p(W)$ equation of state[29] was fitted to the calculated values of $E(W)$ to find the theoretical equilibrium volume $W_{teor} = 1.002W_{exp}$, the bulk modulus $K$ and its pressure derivative $K'$ (see Table 1 below) The lattice parameters $a$, $c$ and $c/a$ determined in this work are presented in Fig. 3, where the experimental results by Vogt et al.[5] are also plotted for comparison.

Fig. 3. Volume dependence of the lattice parameters (a) and $c/a$ ratio (b) of MgB$_2$. The $c/a$ value measured in this work is denoted with solid circle. The open circles and the crosses represent the results of our calculations and the experimental data[5], respectively.

## 5. DISCUSSION

Table 1 presents the MgB$_2$ lattice parameters determined in this work, as well as those measured by various groups[2,5,30,31] and calculated from first principles[3,4,5,6]. As is evident from Table 1, almost all the experimental and theoretical lattice parameters, $a$ and $c$, are in very good agreement with each other (note the coincidence of our theoretical lattice parameters with those calculated in Ref. [4] using the same WIEN97 code). However, the more sensitive $c/a$ ratio obtained in all the calculations is systematically slightly higher than $c/a$ measured experimentally. Among the experimental lattice parameters, the lowest $a$, $c$ and the highest $c/a$ are observed for our samples synthesized at a constant high pressure in the range 6 – 8 GPa. A noticeable spread in lattice parameters measured by various groups for samples obtained at different conditions may be ascribed to an appreciable non-stoichiometry, namely, a possible deficit either in Mg or in B. Notice that the ambient-pressure synthesis at $Т \sim 900 - 950°C$ (see, e.g. Ref. [2]) might lead to the essential sublimation of magnesium (at $p = 0$, $T_{melt}$ of Mg is equal to 650°C), with the resultant deficit in Mg. On the other hand, according to Pearson[32], a deficit in B is practically unavoidable in MgB$_2$. It should be emphasized, however, that at the high-pressure synthesis of MgB$_2$, the sublimation of either element is impossible, because (i) $T_{melt}$ and $T_{subl}$ of parent substances increase with increasing pressure[13] and (ii) the reaction proceeds in closed space.

So, it can be expected that the high-pressure synthesis favors the formation of stoichiometric compound $MgB_2$ with the Mg/B ratio rather close to 1/2. However, on keeping at ambient conditions for a few days, the samples of $MgB_2$ turn out unstable, their color changes, and the intensity of the MgO peaks in the X-ray patterns increases as a result of oxidation. Noteworthy is the study by Jorgensen et al.[30] who applied the refinement of their neutron diffraction data to evaluate the degree of non-stoichiometry of their samples (prepared under argon pressure of 50 bar) and found the ideal Mg/$^{11}$B ratio of 1/2. So far, however, no direct method has been used to investigate the stoichiometry of $MgB_2$. The analysis of various experimental and theoretical results on $MgB_2$ implies that the thorough investigation of its superconducting and other physical properties requires either the obtaining of polycrystalline samples with composition controlled to high accuracy or the growth of $MgB_2$ single crystals.

Table 1. Theoretical and experimental information on structural and mechanical properties of $MgB_2$ at normal conditions

| Reference | $a$ (Å) | $c$ (Å) | $c/a$ | $B$ (GPa) | $B'$ |
|---|---|---|---|---|---|
| Experiment | | | | | |
| [2] | 3.140 | 3.520 | 1.121 | | |
| [5] | 3.086 | 3.521 | 1.141 | 151±5 | 4 |
| | 3.048$^a$ | 3.457$^a$ | 1.134$^a$ | | |
| [30] | 3.085 | 3.521 | 1.141 | 146.8 | |
| | 3.082$^b$ | 3.515$^b$ | 1.140$^b$ | | |
| [31] | 3.084 | 3.523 | 1.142 | | |
| Present | 3.075 | 3.519 | 1.144 | | |
| Theory | | | | | |
| Present | 3.075 | 3.527 | 1.147 | 142.6 | 3.46 |
| [4] | 3.075 | 3.527 | 1.147 | 140.1 | 3.93 |
| [6] | 3.065 | 3.519 | 1.148 | | |
| [3] | 3.071 | 3.528 | 1.149 | | |
| [5] | 3.089 | 3.548 | 1.149 | 139 | |

$^a$ $p$ = 8 GPa.
$^b$ $T$ = 37 K.

The superconducting transition temperature measured for our $MgB_2$ samples synthesized at high pressure is $T_c \sim 36.6$ K, which is slightly lower than $T_c \sim 39$ K found by other groups. This corresponds to a decrease in $T_c$ with decreasing volume (increasing pressure), which was observed for $MgB_2$ samples synthesized at normal pressure[31,33].

While the quadrupole frequency determined in our $^{111}$Cd TDPAC measurements increases with decreasing temperature, no temperature dependence of $n_Q$(FWHM) was observed in the $^{11}$B NMR experiments[9]. This fact, together with the large (> 30%) discrepancy between the $n_Q$ values determined by two different ways in Ref.[10], points out that the temperature dependence of $V_{zz}$ cannot be established reliably in NMR experiments.

The calculated values of EFG's at the B è Mg sites of $MgB_2$ are displayed in Table 2. As is seen from the table, $V_{zz}$ on the B site is about one order of magnitude higher than $V_{zz}$ at the Mg site and is opposite in sign. The value of EFG at the B site is in reasonable agreement with EFG's obtained from $^{11}B$ NMR spectra[8,9,10]. As for $V_{zz}$ at the Mg site, no $^{25}Mg$ NMR measurements have been made for $MgB_2$ so far, which could provide relevant experimental information.

Table 2. EFG's $V_{zz}$ ($10^{16} \times V/cm^2$) obtained both from our TDPAC measurements and from the NMR experiments in comparison with our theoretical results

| Site | | Mg | B |
|---|---|---|---|
| Calculation | | –3.2 | 18.5 |
| $^{111}$Cd PAC | $T_{room}$ | | 15.4 |
| | 143 K | | 18.6 |
| | 0 K[a] | | 20.3 |
| $^{11}$B NMR | [8] | | 16.8 |
| | [9] | | 17.0±0.1 |
| | [10] | | 17.0±1 |
| | | | 22.4±3[b] |

[a] Extrapolation to $T = 0$.
[b] Determined from field dependence.

In the process of thermal diffusion, $^{111}$In — a parent isotope for $^{111}$Cd — could replace both the Mg and B atoms in the $MgB_2$ lattice. Indium is isovalent to boron, but its atomic radius is comparable with that of Mg. Considering that the experimental EFG obtained from our $^{111}$Cd TDPAC measurements of $\mathbf{n}_Q$ is close to calculated $V_{zz}$ at the B site, we suppose that in our experiments, $^{111}$In probe falls into the B site of $MgB_2$ lattice. Probably, a broad distribution of quadrupole frequencies observed in the TDPAC spectra reflects a large degree of disorder occurring around the $^{111}$In/$^{111}$Cd probe because of considerable difference in size between B and In/Cd atoms.

To conclude, our high-pressure technique was applied for synthesis of $MgB_2$. We consider this method to be capable of producing $MgB_2$ samples close to stoichiometry. The value of EFG at the B site, obtained in this work both from *ab initio* calculations and from $^{111}$Cd TDPAC experiments, agrees rather well with those determined from $^{11}B$ NMR spectra. In contrast to the NMR experiments, our TDPAC measurements show that $\mathbf{n}_Q$ and, correspondingly, $V_{zz}$ increase as temperature decreases from room temperature to $T = 4$ K.

*Acknowledgements*—We are grateful to E.G. Maksimov for valuable discussions and to A. Werbel for providing the setup for thermal diffusion. The work was supported by the Russian Foundation for Basic Research, Grant 99-02-17897. M.V. Magnitskaya acknowledges partial financial support from International Science and Technology Center under project No. 207.

# REFERENCES

[1] J. Nagamatsu, N. Nakagawa, T. Muranaka, Y. Zenitani, J. Akimitsu, Nature **410** (2001) 63.

[2] S.L. Bud'ko, G. Lapertot, C. Petrovic, C.E. Cunningham, N. Anderson, P.C. Canfield, Phys. Rev. Lett. **86**, 2001, 1877.

[3] J. Kortus, I.I. Mazin, K.D. Belashchenko, V.P. Antropov, L.L. Boyer, Phys. Rev. Lett. 2001 (in press); cond-mat/0101446.

[4] I. Loa, K. Syassen, Solid State Commun., 2001 (in press); cond-mat/0102462.

[5] T. Vogt, G. Schneider, J.A. Hriljac, G. Hriljac, J.S. Abell, cond-mat/0102480.

[6] J.B. Neaton, A. Perali, cond-mat/0104098.

[7] H. Kotegawa, K. Ishida, Y. Kitaoka, T. Muranaka, J. Akimitsu, cond-mat/0102334.

[8] A. Gerashenko, K. Mikhalev, S. Verkhovskii, cond-mat/0102421.

[9] J.K. Jung, Seung Ho Baek, F. Borsa, S.L. Bud'ko, G. Lapertot, P.C. Canfield, cond-mat/0103040.

[10] H. Tou, H. Ikejiri, Y. Maniwa, T. Ito, T. Takenobu, K. Prassides, Y. Iwasa, cond-mat/0103484.

[11] B.A. Komissarova, G.K. Ryasny, A.A. Sorokin, L.G. Shpinkova, A.V. Tsvyashchenko, L.N. Fomicheva, Phys. Status Solidi B **213**, 1999, 71.

[12] L.G. Khvostantsev, L.F. Vereshchagin, A.P. Novikov, High Temp. High Press. **9**, 1977, 637.

[13] J.K. Kennedy, R. Newton, in: Solids under Pressure (Ed. W. Paul, D.M. Warschauer) McGraw-Hill, New York, 1963.

[14] I. Felner, cond-mat/0102508; D.P. Young, P.W. Adams, J.Y. Chan, F.R. Fronczek, cond-mat/0104063.

[15] J.S. Slusky, N. Rogado, K.A. Regan, M.A. Hayward, P. Khalifah, T. He, K. Inumaru, S. Loureiro, M.K. Haas, H.W. Zandbergen, R.J. Cava, Nature, **410**, 2001, 343.

[16] Y.G. Zhao, X.P. Zhang, P.T. Qiao, H.T. Zhang, S.L. Jia, B.S. Cao, M.H. Zhu, Z.H. Han, X.L. Wang, B.L. Gu, cond-mat/0103077.

[17] Jai Seok Ahn, Eun Jip Choi, cond-mat/0103169; Shao-ying Zhang, Jian Zhang, Tong-yun Zhao, Chuan-bing Rong, Bao-gen Shen, Zhao-hua Cheng, cond-mat/0103203; T. Takenobu, T. Ito, D.H. Chi, K. Prassides, Y. Iwasa, cond-mat/0103241.

[18] S. Sanfilippo, H. Elsinger, M. Nuòes-Regueiro, O. Laborde, S. LeFloch, M. Affronte, G.L. Olcese, A. Palenzona, Phys. Rev. B **61**, 2000, R3800.

[19] A.F. Wells, Structural Inorganic Chemistry, Oxford University Press, Oxford, 1982.

[20] D.V. Filossofov, N.A. Lebedev, A.F. Novgorodov, G.D. Bonchev, G.Ya. Starodub, Preprint JINR Ð6-99-282, Dubna, 1999.

[21] P. Herzog, K. Freitag, M. Reuschenbach, H. Walitzki, Z. Phys. A **294**, 1980, 13.

[22] O.I. Kochetov, A.V. Tsvyashchenko, Ya.P. Bilyalov, A.V. Zernov, A.V. Salamatin, I. Stekl, L.N. Fomicheva, Abst. 5-th Intern. Conf. Nuclear Spectroscopic Investigations of Hyperfine Interactions (NSI-HFI-5), Dubna, 22–24 Sept. 1993, Moscow State Univ. "PRINT" firm, Moscow, 1993, p. 185.

[23] J. Christiansen, P. Heuber, R. Keitel, W. Klinger, W. Loeffler, W. Sandner, W.Witthuhn, Z. Phys. B **24**, 1976, 177.

[24] H. Frauenfelder, R.M. Steffen, in: Alpha-, Beta-, and Gamma-Ray Spectroscopy (Ed. K. Siegbahn), North-Holland, Amsterdam, 1968.

[25] P. Hohenberg, W. Kohn, Phys. Rev. B **136**, 1964, 864; W. Kohn, L. Sham, Phys. Rev. A **140**, 1965, 1133.

[26] P. Blaha, K. Schwarz, J. Luitz, WIEN97, a Full Potential Linearized Augmented Plane Wave Package for Calculating Crystal Properties, Karlheinz Schwarz, Technische Universität Wien, Wien, Austria, 1999, ISBN 3-9501931-0-4.

[27] J.P. Perdew, S. Burke, M. Ernzerhof, Phys. Rev. Lett. **77**, 1996, 3865.

[28] D. Singh, Phys. Rev. B **43**, 1991, 6388.

[29] F. Birch, Geophys. Res. **83**, 1978, 1257.

[30] J.D. Jorgensen, D.G. Hinks, S. Short, cond-mat/0103069.

[31] R. Lorenz, R.L. Meng, C.W. Chu, cond-mat/0102264.

[32] W.B. Pearson, The Crystal Chemistry and Physics of Metals and Alloys, Wiley, New York, 1972.

[33] T. Tomita, J.J. Hamlin, J.S. Schilling, D.G. Hinks, J.D. Jorgensen, cond-mat/0103358.